\documentclass[11pt,a4paper,nofootinbib,superscriptaddress]{revtex4-2}
\usepackage{graphicx,float,wrapfig,subfigure}
\usepackage{amsfonts,amsmath,amssymb,amstext}
\usepackage{latexsym}
\usepackage{bm}
\usepackage{color}
\usepackage{xcolor}
\usepackage[normalem]{ulem}
\usepackage{MnSymbol}
\usepackage[colorlinks=true,linktocpage=true,linkcolor=blue,citecolor=blue]{hyperref}
\usepackage{slashed}
\usepackage{braket}
\usepackage{orcidlink}   
\usepackage{cancel}

\def\be {\begin{equation}}
\def\ee {\end{equation}}
\def\bea {\begin{eqnarray}}
\def\eea {\end{eqnarray}}

\begin{document}
\title{Quarkonium in a QCD medium with momentum-dependent relaxation time}

\author{Sunny Kumar Singh\,\orcidlink{0009-0008-4805-2762}}
\email{sunny.singh@iitgn.ac.in}
\affiliation{Indian Institute of Technology Gandhinagar,Gandhinagar-382355, Gujarat, India}

\author{Samapan Bhadury\,\orcidlink{0000-0002-8693-4482}}
\email{bhadury.samapan@gmail.com}
\affiliation{
Institute of Theoretical Physics, Jagiellonian University ul. St. Lojasiewicza 11, 30-348 Krakow, Poland}

	\author{Ritesh Ghosh\,\orcidlink{0000-0002-6740-7038}}
\email{Ritesh.Ghosh@asu.edu}
\affiliation{College of Integrative Sciences and Arts, Arizona State University, Mesa, Arizona 85212, USA}

	\author{Manu Kurian\,\orcidlink{0000-0001-5667-3333}}
\email{manukurian@iitism.ac.in}
\affiliation{Indian Institute of Technology (Indian School of Mines) Dhanbad, Jharkhand 826004, India}

\begin{abstract}
    In this study, we explore the properties of quarkonia in a hot QCD medium using a newly proposed collision kernel that consistently incorporates the particle's momentum dependence into the relaxation time scale of the medium. The longitudinal component of the gluon self-energy, along with the Debye screening mass, is computed within the one-loop hard thermal loop framework by incorporating non-equilibrium corrections. A modified kinetic theory with an extended relaxation time approximation is employed to model the non-equilibrium dynamics of the QCD medium. The sensitivity of the heavy quarkonia potential to the momentum dependence of the relaxation time is studied. Further, we studied the binding energy and thermal width of quarkonia states within this new kinetic theory. Sizable variations in the temperature behavior of these quantities are observed in comparison with the standard relaxation time approximation method due to the particle momentum dependence on the relaxation timescale of the QCD medium.  Our findings highlight that accounting for the microscopic nature of the collision timescale is crucial for understanding the quarkonium behavior in a QCD medium.
\end{abstract}
	
\maketitle 
	
%\newpage
%\tableofcontents
%--------------------------------
\section{Introduction}
\label{sec:intro}
%--------------------------------
Experimental programs at the Relativistic Heavy Ion Collider (RHIC) and the Large Hadron Collider (LHC) are considered to be the primary laboratories for studying strongly interacting nuclear matter, under extreme conditions. Significant theoretical efforts have been dedicated to understanding the properties of the new phase of matter, Quark-Gluon Plasma (QGP). One of the notable breakthroughs is the success of relativistic dissipative hydrodynamics for describing the collective behavior of the QGP~\cite{Gale:2013da, Heinz:2013th}. The key properties of this QCD natter can be extracted from its collective flow and final particle spectra~\cite{Romatschke:2007mq, Heffernan:2023gye}. In addition, heavy quarks and quarkonia constitute another class of probes that can offer tomographic information about the medium~\cite{vanHees:2005wb,Das:2009vy,Song:2015sfa, Kurian:2019nna, Cao:2018ews,Ghosh:2023ghi,Debnath:2023zet,Carrington:2022bnv,Prakash:2023wbs}. These {\it hard probes} are created in the initial stages of the collision, propagate through the evolving nuclear medium, and interact with it, allowing them to witness the entire QGP evolution~\cite{Song:2020tfm,Kurian:2020orp,Das:2013kea,Das:2016cwd,Singh:2023smw,JETSCAPE:2022hcb}.  

Heavy quarkonia are bound states of heavy quark-antiquark pairs, and their suppression could act as an unambiguous signal for the existence of the QGP in the collision events. The pioneering work of Karsch, Mehr, and Satz laid the theoretical foundation for studying the properties of quarkonium in a QCD medium at a finite-temperature by utilizing potential models~\cite{Karsch:1987pv}. Further, Matsui and Satz have demonstrated that the color screening in the QGP could lead to the dissociation of quarkonia~\cite{Matsui:1986dk}. Recent theoretical studies have explored the in-medium modifications of the heavy quark–antiquark potential, paving the way for a more realistic description of quarkonium suppression in the heavy-ion collision experiments~\cite{Dumitru:2009ni,Young:2011ug,Margotta:2011ta,Escobedo:2011ie,Riek:2010py,Thakur:2012eb,Rothkopf:2011db,Burnier:2007qm,Mocsy:2007jz,Bonati:2015dka,Jamal:2018mog,Sebastian:2023tlw,Singh:2023zxu,Debnath:2023dhs,Debnath:2025cmd,Cabrera:2006wh,Guo:2018vwy,Agotiya:2008ie}. These modifications to the quarkonium potential can be introduced through the dielectric properties of the underlying medium.

The medium created at the collision experiments at the RHIC and LHC is not instantaneously thermalized. Instead, it undergoes a pre-equilibrium phase, which is characterized by momentum-space anisotropies and rapid longitudinal expansion. Such pre-equilibrium conditions can further the quarkonium potential and thereby its properties in the evolving medium. This sets the motivation for the present study. A comprehensive understanding of how non-equilibrium effects influence heavy quarkonium requires the proper knowledge of the system away from equilibrium. One natural way to model the non-equilibrium contribution is by solving kinetic theory. The most standard approach for this is the Anderson–Witting relaxation time approximation (RTA)~\cite{Anderson:1974nyl}. However, the traditional RTA framework does not fully respect the conservation laws in relativistic kinetic theory if the relaxation time becomes momentum dependent~\cite{Jaiswal:2016sfw, Dash:2023ppc,Singh:2024leo}. To address this, novel approaches have been introduced recently~\cite{Mitra:2020gdk, Rocha:2021zcw, Dash:2021ibx}, which offer improved theoretical consistency.

In the current analysis, we utilize the recently developed extended relaxation time approximation (ERTA)~\cite{Dash:2021ibx, Dash:2023ppc} to model the non-equilibrium corrections to the quarkonium properties. We begin by evaluating the non-equilibrium corrections to the longitudinal component of the gluon self-energy and the Debye screening mass within the framework of one-loop hard thermal loop (HTL) theory~\cite{Bellac:2011kqa,Haque:2024gva}. These corrections form the basis for a modified heavy quarkonium potential in an evolving QGP medium. We then investigate the in-medium behavior of quarkonium states, focusing specifically on their binding energy and thermal width while accounting for non-equilibrium effects. Further, we perform a direct comparison of the ERTA  corrections with the conventional RTA observations with a parameterized momentum-dependent form of the thermal relaxation time. 

The manuscript is organized as follows. In Section~\ref{II}, we present the formalism of the present analysis. Subsection~\ref{II.1} discusses the contributions from gluon and quark loops to the heavy quarkonium potential in the evolving QGP. We reviewed the ERTA framework in Subsection~\ref{SecB}. Subsection~\ref{SecC} outlines the modified screening mass within the ERTA, while Subsection~\ref{II.3} describes the non-equilibrium modifications to the quarkonium binding energy and thermal width in the medium. The results and corresponding discussion are presented in Section~\ref{III}. Finally, in Section~\ref{IV}, we summarize this study with an outlook.

\textit{Notations and conventions}: Throughout this paper, we use the metric $g^{\mu\nu} = \mathrm{diag}(1,-1,-1,-1)$. A four-vector is denoted by $A^\mu\equiv(a^0,\bf{a})$, with $A^2 = A_\mu A^\mu$. A lowercase letter $a=|\bf{a}|$ denotes the magnitude of a particle’s three-momentum $\mathbf{a}$.

%%%%%%%%%%%%%%%%%%%%%%%%%%%%%%%%%%%%%%%%%%%%%%%%%%%%%%%%%%%%%%%
\section{Formalism}\label{II}
%%%%%%%%%%%%%%%%%%%%%%%%%%%%%%%%%%%%%%%%%%%%%%%%%%%%%%%%%%%%%%%
\subsection{Quarkonium potential in QCD medium}\label{II.1}
%%%%%%%%%%%%%%%%%%%%%%%%%%%%%%%%%%%%%%%%%%%%%%%%%%%%%%%%%%%%%%%
Understanding the dissociation of quarkonium states near the QCD crossover temperature requires a realistic treatment of the heavy quark-antiquark potential that includes non-perturbative effects. While the Cornell potential, comprising of a short-range Coulomb interaction and a long-range linear confinement term, accurately captures quarkonium properties in vacuum, its extension to a thermal medium is far more complex and remains less well established. To model the medium modifications, we employ a framework in which the in-medium potential in coordinate space is derived through a Fourier transform of the momentum-space potential, modified by the medium’s dielectric response. This potential can be expressed as~\cite{Lafferty:2019jpr,Thakur:2020ifi}:
\bea
V(r) = \int \frac{d^3p}{(2\pi)^{3}} \left(e^{i\boldsymbol{p} \cdot \boldsymbol{r}} - 1\right) \frac{V_{\text{Cornell}}(p)}{\epsilon(p)}, \label{HQ_Vr} 
\eea where $\epsilon(p)$ encodes the screening effects of the thermal medium. The momentum-space form of the Cornell potential takes the form as,
\bea V_{\text{Cornell}}(p) = - \frac{4\pi\alpha}{p^2} - \frac{8\pi\sigma}{ p^4}.
\eea
Here, $\alpha=C_F \alpha_s$ with $\alpha_s$ as the running coupling constant and $C_F=(N_c^2-1)/2N_c$ with $N_c=3$ as the number of color charges, and
$\sigma$ represents the string tension responsible for quark confinement. The dielectric permittivity $\epsilon(p)$ plays a crucial role in describing medium effects on the heavy-quark potential. It is defined as~\cite{Thakur:2013nia,Thakur:2020ifi}: 
\begin{equation} 
    \epsilon^{-1}(p) = \lim_{p_0 \to 0} p^2  \bar{D}_{11}(P), 
    \label{eq:eps_def}
\end{equation}
where $ \bar{D}_{11}(P)$ denotes the longitudinal part of the `11'-component of the resummed gluon propagator in the real-time formalism. This relationship reflects how medium-induced screening alters the interaction between a heavy quark and antiquark ultimately influencing the stability and dissociation behavior of quarkonium states. The propagator $\bar{D}_{11}(P)$ can be decomposed in terms of the retarded, advanced, and symmetric resummed propagators as~\cite{Dumitru:2009fy,Du:2016wdx}: 
%%%%
\begin{equation}
\bar{D}_{11}(P) = \frac{1}{2} \left( \bar{D}_R(P) + \bar{D}_A(P) + \bar{D}_F(P) \right). \label{eq:D11_decomp} 
\end{equation} 
%%%%
Using this decomposition along with the relevant expressions for the propagators, the inverse dielectric function of an expanding QCD medium can be written as: 
%%%
\begin{equation} 
\epsilon^{-1}(p) = \frac{p^2}{p^2 + m_{D}^2} - i \frac{\pi T p m_{D}^2}{(p^2 + m_{D}^2)^2}, 
\label{eq:eps_bulk} 
\end{equation}
%%%
where  $m_{D}$ is the Debye screening mass.
The first term in Eq.~\eqref{eq:eps_bulk} represents the medium-induced screening, while the second (imaginary) term accounts for the Landau damping.
Note that the real part of the screened quarkonia potential is obtained from the Fourier transform of $\frac{1}{2} \left( \bar{D}_R + \bar{D}_A\right)$. In contrast, the imaginary component, associated with Landau damping, originates from the Fourier transform of the symmetric part of the gluon propagator $\bar{D}_F$.
Using Eq.~{\eqref{eq:eps_bulk}} in Eq.~\eqref{HQ_Vr}, the screened (real) potential between heavy quarks in the medium becomes,
\begin{align}
\text{Re\,} V(r)& = -\int \frac{d^3p}{(2\pi)^{3}} \left(e^{i\boldsymbol{p} \cdot \boldsymbol{r}} - 1\right)\left(\frac{4\pi\alpha}{p^2} + \frac{8\pi\sigma}{ p^4}\right) \frac{p^2}{p^2 + m_{D}^2} \nonumber\\
&=-\alpha\, m_{D}\left(\frac{e^{-\hat r}}{\hat r }+1\right)+\frac{2\sigma}{ m_{D} }\left(1+\frac{e^{-\hat r }-1}{\hat r }\right),
\end{align}
with $\hat r=r \; m_{D}$ and the imaginary part of the potential takes the form as,
\bea
\text{Im\,} V(r) = \int \frac{d^3p}{(2\pi)^{3}} \left(e^{i\boldsymbol{p} \cdot \boldsymbol{r}} - 1\right)\left(\frac{4\pi\alpha}{p^2} + \frac{8\pi\sigma}{ p^4}\right) \frac{\pi T p , m_{D}^2}{(p^2 + m_{D}^2)^2}.
\label{eq:ImV}
\eea
Defining the following integrals,
\begin{subequations}
    \begin{align}
        %1
        \chi(x)=2\int^\infty_0 \frac{dz}{z(z^2+1)^2}\Bigg[1-\frac{\sin xz}{xz}\Bigg],\\
        %2
        \phi_n(x)=2\int^\infty_0 \frac{dz\; z}{(z^2+1)^n}\Bigg[1-\frac{\sin xz}{xz}\Bigg],
    \end{align}
\end{subequations}
we can write the imaginary part of the heavy quarkonia potential as,
\begin{align}
    \mathrm{Im} \; V(r)&=-T\alpha \phi_2 (\hat{r}) - 2\frac{T\sigma}{m_D^2}\chi(\hat{r}), \nonumber\\
    &= \mathrm{Im}\, V_{\text{Coulomb}}(\hat{r}) + \mathrm{Im}\, V_{\text{String}}(\hat{r}).
    \label{eq:ImV1}
\end{align}
 In the Eq.~(\ref{eq:ImV1}) above, the first term is due to the Coulomb part of the Cornell potential, and the second term is called the String part of the imaginary potential. We note that the Debye screening mass plays a central role in determining both the real and imaginary parts of the in-medium quarkonia potential. Now, we proceed to calculate the Debye mass in the relevant thermal medium.

\subsubsection*{Gluon self energy and screening mass}\label{II.2}

In a thermal medium, the propagation of gluons is modified by the presence of thermal quarks and gluons, leading to the screening of color charges. This effect is quantified by the Debye mass, which governs the exponential suppression of long-range chromoelectric fields. The non-equilibrium effects due to the expansion of the medium can be captured by the screening mass. To consistently incorporate both equilibrium and non-equilibrium dynamics, we evaluate the self-energies and propagators using the well-known Keldysh real-time formalism~\cite{Carrington:1997sq, Mrowczynski:2000ed}, which is particularly suited for studying time-dependent phenomena in a thermal field theory, as discussed in Refs.~\cite{Chou:1984es,Carrington:1998jj,Dumitru:2009fy,Mrowczynski:2016etf,Nopoush:2017zbu}. Within this approach, the longitudinal component of the gluon self-energy receives contributions from gluon and quark loop diagrams. The longitudinal component of gluon self-energy with quark-loop is given as~\cite{Mrowczynski:2000ed,Dumitru:2009fy,Nopoush:2017zbu},
\begin{align}
    \Pi^L_{q,R}(P) &=i 2 N_f g_s^2\int \frac{d^4K}{(2\pi)^4}(q_0 k_0+\mathbf{q}\cdot \mathbf{k}+m^2)\frac{2\pi i \,\left(1-2 f(\mathbf{k})\right)\delta(K^2-m^2)}{Q^2-m^2-i\,\text{sgn}(q_0)\,\epsilon},
    \label{eq: piR1}
\end{align}
where $Q^\mu= K^\mu - P^\mu$, $g_s$ is the coupling constant, $N_f$ is the number of quark flavors and $d^4K$ is the momentum integration measure. The distribution function, $f(\textbf{k})$ contains the details of the hydrodynamically evolving medium. Carrying out the $k_0$ integration using the pole structure of the propagators and applying the Hard Thermal Loop (HTL) approximation, the expression can be simplified considerably. After angular integration and keeping only the leading-order HTL contributions, one finally obtains,
\begin{align}\label{massive}
\Pi_{q,R}^L(P)
    =&\frac{N_f g_s^2}{4\pi^3}\int\frac{k^2 dk\,d\Omega}{E_k}f(k)\frac{\frac{E_k^2}{k^2}-(\hat{\mathbf{k}} \cdot \hat{\mathbf{p}})^2}{\left(\hat{\mathbf{k}} \cdot \hat{\mathbf{p}}+\frac{p_0+i\epsilon }{p\,k} E_k\right)^2},
\end{align}
where we consider the number of quark flavors as $N_f=3$.   
The quark contribution to the Debye mass in QCD is obtained as,
\begin{align}
    m_{D,q}^2=&-\Pi_R^L(p_0=0,p\rightarrow 0, m).
\end{align}
The gluon and ghost loop contributions to the Debye screening mass in an evolving QCD medium can be expressed within the HTL as,
\begin{eqnarray}
    m^2_{D,g}=\frac{2C_A g_s^2}{\pi^2}\int_0^\infty \,k\,dk \, f,
\end{eqnarray}
where $C_A$ is the group factor~\cite{Bellac:2011kqa}. For the system out of equilibrium, we have $f=f_0+\delta f$. In equilibrium, the screening properties are governed by the equilibrium distribution function $f_0$. However, in a system slightly away from equilibrium, such as the QGP produced in heavy-ion collisions, the distribution function acquires a non-equilibrium correction $\delta f$, reflecting the response of the system to external perturbations. The microscopic dynamics of the medium determine the precise form and encode essential information about its relaxation properties. One approach to model this is through kinetic theory.  Below, we discuss the formulation of relativistic hydrodynamics from kinetic theory, based on the ERTA.

\subsection{Relativistic hydrodynamics with ERTA}\label{SecB}

Before we obtain the expressions of the Debye mass and hence the potential, we would like to give a brief overview of the framework of ERTA hydrodynamics, based on which the non-equilibrium corrections are calculated.

%~~~~~~~~~~~~~~~~~~~~~~~~~~~~~~~~~~~~~~~~~~~~~~~~~~~~~~~~~
\subsubsection{Relativistic hydrodynamics}\label{sssec:RH}
%~~~~~~~~~~~~~~~~~~~~~~~~~~~~~~~~~~~~~~~~~~~~~~~~~~~~~~~~~

As the system under consideration is made of massless and chargeless particles, the only relevant conserved current is the energy-momentum tensor (associated with the translational invariance). Under the choice of Landau frame ($U_\nu T^{\mu\nu} = \varepsilon U^\mu$) the conserved energy-momentum tensor can be written as,
\begin{eqnarray}
        T^{\mu\nu} = \varepsilon\, U^\mu U^\nu - \mathcal{P} \Delta^{\mu\nu} + \pi^{\mu\nu} = \int \frac{d^3 p}{\left(2\pi\right)^3 p} P^\mu P^\nu f (X,P), \label{T^mn-def}
\end{eqnarray}
where $\varepsilon$ is the out-of-equilibrium energy density, $\mathcal{P}$ is the isotropic pressure, $\pi^{\mu\nu}$ is the shear viscous pressure, and $\Delta^{\mu\nu} = g^{\mu\nu} - U^\mu U^\nu$ is the projection operator orthogonal to the fluid four-velocity $U^\mu$. Additionally, the Landau matching choice requires that, $\varepsilon$ be equal to the equilibrium energy density ($\varepsilon_0$), i.e., $\varepsilon = \varepsilon_0$. We can then express the macroscopic variables, $\varepsilon_0$, $\mathcal{P}$ and $\pi^{\mu\nu}$ as,
\begin{subequations}
    \begin{align}
        %1
        \varepsilon_0 &= U_\mu U_\nu T^{\mu\nu} = \int \frac{d^3 p}{\left(2\pi\right)^3 p} \left(U\cdot P\right)^2 f_0 (X,P), \label{vare_0} \\
        %2
        \mathcal{P} &= - \frac{1}{3} \Delta_{\mu\nu} T^{\mu\nu} = - \frac{1}{3} \int \frac{d^3 p}{\left(2\pi\right)^3 p} \left(P\cdot \Delta \cdot P\right) f_0 (X,P), \label{P} \\
        %3
        \pi^{\mu\nu} &= \Delta^{\mu\nu}_{\alpha\beta} T^{\alpha\beta} = \Delta^{\mu\nu}_{\alpha\beta} \int \frac{d^3 p}{\left(2\pi\right)^3 p} P^{\alpha} P^{\beta} f (X,P), \label{pi^mn}
    \end{align}
\end{subequations}
where $\Delta^{\mu\nu}_{\alpha\beta} = \frac{1}{2} \left(\Delta^\mu_\alpha \Delta^\nu_\beta + \Delta^\mu_\beta \Delta^\nu_\alpha\right) - \frac{1}{3} \Delta_{\alpha\beta} \Delta^{\mu\nu}$ is the doubly symmetric traceless projection operator that is orthogonal to fluid four-velocity.
The conservation law for energy-momentum tensor ($\partial_\mu T^{\mu\nu} = 0$) leads to the evolution equation for $\varepsilon_0$ and $U_\mu$ as,
\begin{subequations}
    \begin{align}
        %1
        \dot\varepsilon_0 + (\varepsilon_0+\mathcal{P})\theta - \pi^{\mu\nu} \sigma_{\mu\nu} &= 0,  \label{heq1}\\
        %2
        (\varepsilon_0+\mathcal{P})\dot U^\alpha - \left(\nabla^\alpha \mathcal{P}\right) + \Delta^\alpha_\nu \partial_\mu \pi^{\mu\nu}  &= 0, \label{heq2}
    \end{align}
\end{subequations}
with $\theta \equiv \partial \cdot U$ being the expansion, and $\sigma^{\mu\nu} = \Delta_{\alpha\beta}^{\mu\nu} \left(\partial^\alpha U^\beta\right)$ being the shear velocity tensor. The spacetime derivative has been split into components along and orthogonal to $U_\mu$, yielding $\partial_\mu = U_\mu D + \nabla_\mu$, where $D \equiv \left(U\cdot \partial\right)$ is the co-moving derivative and $\nabla_\mu \equiv \Delta_\mu^\alpha \partial_\alpha$ is the spacelike derivative. We have also used the shorthand notation, $\Dot{A} = D A = U^\mu \partial_\mu A$ for the co-moving derivative.
Using the definitions from Eqs.~\eqref{vare_0}-\eqref{pi^mn} into Eqs.~\eqref{heq1}-\eqref{heq2}, we obtain the evolution equations for temperature and fluid four-velocity as \cite{Dash:2021ibx},
\begin{align}
    \Dot{\beta} = \frac{\beta}{3} \theta + \mathcal{O} \left(\partial^2\right),
    \qquad{\rm and,}\qquad
    \beta \Dot{U}_\mu = - \nabla_\mu \beta + \mathcal{O} \left(\partial^2\right), \label{beta,u_evol}
\end{align}
where $\mathcal{O} \left(\partial^2\right)$ represents corrections from second and higher orders that have been ignored presently. The relations of Eq.~\eqref{beta,u_evol} will be used later, to derive the non-equilibrium correction, $\delta f$.

%~~~~~~~~~~~~~~~~~~~~~~~~~~~~~~~~~~~~~~~~~~~~~~~~
\subsubsection{The ERTA framework}\label{SsecB.2}
%~~~~~~~~~~~~~~~~~~~~~~~~~~~~~~~~~~~~~~~~~~~~~~~~

It has been known for a long time that for a gas of classical particles, the particles with lower momenta attain the equilibrium distribution at an earlier time than those with higher momenta \cite{krook1976formation}. Thus, it is natural to presume that the relaxation time for particles of different momenta will be different as well. Recent studies \cite{Dusling:2009df, Teaney:2013gca} have also arrived at similar conclusions while considering the dynamics of the QCD system. These studies then inspired the development of several new kinetic theoretical frameworks where the relaxation time was allowed to be dependent on momenta \cite{Rocha:2021zcw, Dash:2021ibx, Shaikh:2024gsm, Biswas:2022cla}. In the present analysis, we employ the ERTA framework to determine the out-of-equilibrium distribution function up to first order in spacetime gradient. Below, we provide a brief overview of the first-order ERTA approach.

The relativistic Boltzmann equation, in the absence of any external force, under the RTA takes the form as \cite{Anderson:1974nyl},
\begin{align}
    P^\mu \partial_\mu f = - \frac{\left(U\cdot P\right)}{\tau_{\rm R}} \left(f - f_{0}\right), \label{Beq-RTA}
\end{align}
where the equilibrium distribution function $f_0$ is given by,
\begin{align}
    f_0 (X,P) = \frac{1}{\exp\big[\beta \left(U\cdot P\right) + \alpha\big] + a}, \label{f_0-def}
\end{align}
with $\beta = 1/T$ being the inverse temperature, $U_\mu$ as the fluid four-velocity, the dimensionless quantity $\alpha = \mu/T$ being the ratio of chemical potential to temperature, and $a$ taking the values $0, 1$, and $ -1$ for particles following Maxwell-Boltzmann (MB), Fermi-Dirac (FD), and Bose-Einstein (BE) statistics, respectively.

For a system in an out-of-equilibrium state, the phase-space distribution function can be split into equilibrium and non-equilibrium parts by following the Chapman-Enskog expansion as,
\begin{align}
    f(X,P) = f_0(X,P) + \delta f(X,P), \label{ooe_f}
\end{align}
where the condition $\frac{\delta f}{f_0} \ll 1$ is imposed for the formulation of hydrodynamic theories. In such systems, the notions of temperature, chemical potential, and fluid velocity are ill-defined, and hence $T, \mu$ and $U_\mu$ correspond to some auxiliary variables that coincide with respective variables only at equilibrium. Consequently, the equilibrium state ($f_0$) used in Eq.~\eqref{ooe_f} is an auxiliary equilibrium \cite{Dash:2021ibx}. The out-of-equilibrium variables, $T,\,\,\mu$, and $U_\mu$ (which we call \textit{hydrodynamic variables} from this point onward) are fixed only through frame and matching conditions. However, in relativistic hydrodynamics, these frame and matching conditions are completely arbitrary, and hence the definitions of hydrodynamic variables are also a matter of choice. As a result of this ambiguity, one may expect that the equilibrium distribution function as described in Eq.~\eqref{f_0-def} is not the actual equilibrium state of the system, as mentioned above. However, in the case of the RTA, one is forced to choose the Landau frame and matching conditions to ensure conservation law, as a result of which the hydrodynamic variables end up being the variables associated with the equilibrium state, provided the relaxation time is momentum-independent. When the relaxation time is momentum-dependent, the conservation laws can no longer be guaranteed, and this led to the proposal of the ERTA, which modifies the Boltzmann equation as,
\begin{align}
    P^\mu \partial_\mu f = - \frac{\left(U\cdot P\right)}{\tau_{\rm R}} \left(f - f_{0}^*\right), \label{Beq-ERTA}
\end{align}
where $f_{0}^*$ is given by,
\begin{align}
    f_0^* (X,P) = \frac{1}{\exp\big[\beta^* \left(U^*\cdot P\right) + \alpha^*\big] + a}, \label{f*_0-def}
\end{align}
with $\beta^* = 1/T^*$, $\alpha^* = \mu^*/T^*$. Here, $T^*, \mu^*$, and $U_\mu^*$ correspond to the temperature, chemical potential, and the fluid four velocity of the system at equilibrium. The starred quantities are called the \textit{thermodynamic variables}, and consequently, $f_0^*$ represents the true thermodynamic local equilibrium of the system toward which the out-of-equilibrium system relaxes as $t\to \infty$. It should be noted that in general, the variables, $\beta^*, U_\mu^*$ and $\mu^*$ are functions of spacetime. The advantages of this collision kernel in Eq.~\eqref{Beq-ERTA} are that (i) the relaxation time can now be taken to be momentum-dependent \cite{Dash:2021ibx}, and (ii) one is free to choose any arbitrary frame and matching condition \cite{Bhadury:2024ckc, Mukherjee:2025dqp}. The thermodynamic variables and hydrodynamic variables are related to each other through the relation,
\begin{align}
    U_\mu^* = U_\mu + \delta U_\mu,
    \qquad
    T^* = T + \delta T,
    \qquad
    \mu^* = \mu + \delta \mu, \label{therm-hyd_rel}
\end{align}
where the correction terms (denoted with $\delta$ in front) can be determined through frame and matching conditions \cite{Dash:2021ibx}. Since the ERTA provides us with the expressions of the transport coefficients that are invariant under choice of frame and matching conditions, for the present work, we choose the Landau-Lifshitz version \cite{landau1987fluid} of the frame and matching condition. Consequently, following the results of Ref.~\cite{Dash:2021ibx}, we can write, 
\begin{align}
    \delta U_\mu = \beta \mathcal{C}_1 \left(\nabla_\mu \alpha\right),
    \qquad
    \delta T = \mathcal{C}_2\, \theta,
    \qquad
    \delta \mu = \mathcal{C}_3\, \theta, \label{delX-expr}
\end{align}
where the coefficients $\mathcal{C}_{1}, \mathcal{C}_{2}$ and $\mathcal{C}_{3}$ can be found in Refs.~\cite{Dash:2021ibx, KumarSingh:2025kml}. In the present study, the system under consideration consists of chargeless, massless particles and hence, we have $\mathcal{C}_{1} = \mathcal{C}_{2} = \mathcal{C}_{3} = 0$. Note that the shear viscosity coefficient will depend on the nature of the momentum dependence of the relaxation time. In the present work, we will consider the following parameterization of the relaxation time,
\begin{align}
    \tau_{\rm R} (X,P) = t_{\rm R} (X) \left(\frac{U\cdot P}{T}\right)^\ell.\label{tau_R-param}
\end{align}
Here, $t_{\rm R}$ denotes the momentum-independent component, while the parameter $\ell$ characterizes the momentum dependence of the relaxation time\footnote{In this work, we focus on the range $0\leq \ell \leq 1$ which is the region of interest for a QCD medium \cite{Dusling:2009df,Teaney:2013gca}.}. In principle, the parameter $\ell$ can depend on the temperature \cite{Mukherjee:2025dqp}. However, for the present study, we confine our analysis to the case where $\ell$ is independent of temperature. The temperature-dependent scenario will be explored in future.

By solving the relativistic Boltzmann equation as described in Eq. \eqref{Beq-ERTA}, we obtain the non-equilibrium correction as \cite{Dash:2021ibx},
\begin{align}
    \delta f_{(1)} = \frac{\beta \tau_{\rm R} (X,P)}{\left(U\cdot P\right)} P^\mu P^\nu \sigma_{\mu\nu} f_0 \Tilde{f}_0 = \frac{\beta^{1 + \ell} t_{\rm R} (X)}{\left(U\cdot P\right)^{1-\ell}} P^\mu P^\nu \sigma_{\mu\nu} f_0 \Tilde{f}_0, \label{del_f-ERTA}
\end{align}
where the subscript `$(1)$' in $\delta f_{(1)}$ denotes that only the first-order corrections have been considered, and we have defined $\Tilde{f}_0 = 1 -a f_0$. As mentioned before, in the derivation of Eq.~\eqref{del_f-ERTA}, we had to consider the hydrodynamic evolution Eq.~\eqref{beta,u_evol}.

%~~~~~~~~~~~~~~~~~~~~~~~~~~~~~~~~~~~~~~~~~~~~~~~
\subsubsection{Transport coefficients}\label{TC}
%~~~~~~~~~~~~~~~~~~~~~~~~~~~~~~~~~~~~~~~~~~~~~~~

Having determined the non-equilibrium correction to the distribution function in Eq.~\eqref{del_f-ERTA}, we can use Eq.~\eqref{pi^mn} to obtain the sole transport coefficient of fluid under consideration, i.e., the shear viscosity. In Eq.~\eqref{pi^mn} the equilibrium part of the distribution function ($f_0$) gives zero contribution and using Eq.~\eqref{del_f-ERTA} we obtain the shear viscous pressure as,
\begin{align}
    \pi^{\mu\nu} = 2 \beta K_{32} \sigma^{\mu\nu}, \label{pi^mn-result}
\end{align}
where the thermodynamic integral is defined as,
\begin{align}
    K_{nm} \equiv \frac{1}{(2m+1)!!} \!\int \!\frac{d^3 p}{\left(2\pi\right)^3 p} \tau_{\rm R} \left( U\cdot P \right)^{\!n-2m} \!\left(P \cdot \Delta \cdot P\right)^{\!m} f_{0} \Tilde{f}_0. \label{K_nq-def}
\end{align}
This reduces to $t_{\rm R} \beta^{\ell} J_{n+\ell,m}$, if the parametrization of Eq.~\eqref{tau_R-param} is chosen, with the definition,
\begin{align}
    J_{nm} \equiv \frac{1}{(2m+1)!!} \!\int \!\frac{d^3 p}{\left(2\pi\right)^3 p} \left(U \cdot P \right)^{\!n-2m} \!\left(P \cdot \Delta \cdot P\right)^{\!m} f_{0} \Tilde{f}_0. \label{I_nq-def}
\end{align}
Comparing the result of Eq.~\eqref{pi^mn-result} with the Navier-Stokes relation, $\pi^{\mu\nu} = 2 \eta\, \sigma^{\mu\nu}$. We can identify the shear viscosity as \cite{Dash:2021ibx},
\begin{eqnarray}
    \eta = t_\mathrm{R}\beta^{\ell+1} J_{3+\ell,2}. \label{eta}
\end{eqnarray}
In the next part, we discuss how the system under consideration behaves when the fluid undergoes a boost-invariant expansion along a particular direction (say the $z$-axis or the beam axis in heavy-ion collision) with the transverse direction being homogeneous and isotropic \cite{Bjorken:1982qr}.

\subsubsection{Bjorken expansion}
The non-equilibrium correction to the distribution function as given in Eq.~(\ref{del_f-ERTA}) depends on the four-velocity of the fluid element via the velocity stress tensor $\sigma_{\mu\nu}$ as defined below Eq.~\eqref{heq2}. In this work, we choose this velocity profile to be of a 1-D longitudinal boost-invariant expansion given by $U^\mu=(t/\tau,0,0,z/\tau)$ where $\tau=\sqrt{t^2-z^2}$ is the proper time of the fluid element. In terms of Milne coordinates $(\tau,x,y,\eta_s)$, where $\eta_s = \tanh^{-1}(z/t)$ is the space-time rapidity, the fluid velocity becomes $U^\mu=(1,0,0,0)$ and the metric takes the form, $g^{\mu\nu}=(1,-1,-1,-1/\tau^2)$. With this, the temporal derivative becomes  $D=U^\mu\partial_\mu=\partial/\partial \tau$ and the expansion scalar takes the form $\theta=\partial_\mu U^\mu =1/\tau$. We also have $\sigma^{\mu\nu}=\text{diag}\left[0,1/(3\tau),1/(3\tau),-2/(3\tau^3)\right]$ and this leads to $\pi^{\mu\nu}\sigma_{\mu\nu}=4\eta/(3\tau^2)$ in Eq.~(\ref{heq1}) of the energy evolution equation. The energy evolution equation can now be used to derive the evolution of temperature in the fluid element as a function of the proper time,
\begin{align}
\frac{dT}{d\tau} = \frac{1}{d\varepsilon_0/dT}\left[\frac{4\eta}{3\tau^2} -\frac{\varepsilon_0+\mathcal{P}}{\tau}\right],\label{T-evo}
\end{align}
where $\eta$ is given by Eq.~(\ref{eta}) and we employ the conformal equation of state, $\varepsilon_0=3\mathcal{P}$. The temperature as a function of the proper time, $\tau$, can be determined by numerically solving the above equation \eqref{T-evo}, which can be used to study the screening properties of the evolving medium.

Next, we will use the information obtained so far to determine the gluon self-energy, screening mass, and other phenomenologically relevant variables to study the properties of quarkonium states in the fluid system under consideration.

%%%%%%%%%%%%%%%%%%%%%%%%%%%%%%%%%%%%%%%%%%%%%%%%%%%%
\subsection{Modified screening within the ERTA}\label{SecC}
%%%%%%%%%%%%%%%%%%%%%%%%%%%%%%%%%%%%%%%%%%%%%%%%%%%%%%%
The non-equilibrium corrections to the gluon self-energy from quark loop in the massless case can be obtained from Eq.~\eqref{massive} by setting $E_k = k$ as,
\begin{align}
    \delta \Pi_{R}(P) = \frac{N_f g_s^2}{4 \pi^3} \int k \, {dk} \, d\Omega \,  \frac{1-(\hat{\bf k}\cdot \hat{\bf p})^2}{\left(\hat{\bf k}\cdot \hat{\bf p} + \frac{p^0+i\epsilon}{p}\right)^2}\,\delta f .
\end{align}
 Here, we have $\hat{\bf k}\cdot \hat{\bf p} = \cos \theta_{kp}$ and $d\Omega= \sin{\theta_{kp}} d\theta_{kp} d\phi$ where $\theta_{kp}$ is the angle between the particle and gluon momentum. With this, the expression for the gluon self-energy correction becomes,
\begin{align}
    \delta \Pi_R(P) = \frac{N_f g_s^2}{2 \pi^2} \int_0^\infty \int_0^\pi  k\;dk \,  \sin\theta_{kp}\; {d\theta_{kp}}\frac{1- \cos^2 \theta_{kp}}{\left(\cos\theta_{kp} + \frac{p^0+i\epsilon}{p}\right)^2} \delta f \,.
\end{align}
In general, the non-equilibrium correction to the distribution function ($\delta f$) depends on the particle momentum $k$, the angle $\theta_{kp}$, and the hydrodynamic velocity $U^\mu$ of the fluid element. This is apparent when we consider a fluid velocity profile of a purely longitudinal boost-invariant expansion. In this scenario, we can express the first-order correction to the distribution function as derived using momentum-dependent relaxation time within the ERTA framework given in Eq.~(\ref{del_f-ERTA}) as,
\begin{align}\label{shearf}
    &\delta f\equiv \delta f_{(1)} = \frac{\beta \tau_{\rm R}}{\left(U\cdot K\right)} \frac{k^2}{3\tau}(1-3\cos^2 \theta_{kp}) f_{0} \Tilde{f}_{0}.  
\end{align}
A detailed derivation can be found in Appendix~\ref{Appendix:B}. By utilizing the parameterized form of relaxation time as described in Eq. \eqref{tau_R-param} in Eq. \eqref{shearf}, the non-equilibrium contribution of the self-energy can be expressed as, 
\begin{align}
    \delta \Pi_R(P)= \frac{N_f g_s^2}{2\pi^2} \frac{t_\mathrm{R}}{3\tau} \frac{1}{T^{\ell+1}} \int_0^\infty k^{\ell+2} \, {dk} f_0 \Tilde{f}_0  \int_0^\pi \sin{\theta} (1-3\cos^2 \theta) d\theta\frac{\left(1-\cos^2 \theta\right)}{\left(\cos \theta + \frac{p^0+i\epsilon}{p}\right)^2}.
\end{align}
Since the structure of $\delta \Pi_R(P)$ is the same for the gluon loop as well, we end up with,
\begin{align}
    \delta \Pi_R(P) = \frac{g_s^2 t_\mathrm{R} \beta^{\ell+1}}{6\pi^2 \tau} (N_f I_{\ell}^{(q)}+N_c I_{\ell}^{(g)}) \left[I_\theta (1,0)-3 I_\theta (1,2)\right],
\end{align}
where the integrals in the above expression are defined as,
\begin{subequations}
    \begin{align}
        I_\theta (n,m) &= \int^\pi_0 \sin^n \theta \cos^m \theta \frac{1-\cos^2 \theta}{\left(\cos \theta + \frac{p^0+i\epsilon}{p}\right)^2}d\theta,\label{theta-integral} \\
        I_{\ell}^{(q)} &= \int^\infty_0 k^{\ell+2} \frac{e^{k/T}}{\left(e^{k/T}+1\right)^2}dk,\label{q-gluon}\\
        I_{\ell}^{(g)} &= \int^\infty_0 k^{\ell+2} \frac{e^{k/T}}{\left(e^{k/T}-1\right)^2}dk.\label{I-gluon}
    \end{align}
\end{subequations}
While the angular integrals are the same for both quark and gluon loop contributions to the gluon self-energy, the $k$ integrals for quark and gluon contributions, denoted by $I_{\ell}^{(q)}$ and $I_{\ell}^{(g)}$ respectively, must be evaluated separately. The $I_{\ell}^{(q)}$ and $I_{\ell}^{(g)}$ integrals are estimated in Appendix~\ref{Appendix2}. With this, we obtain the gluon self-energy correction as,
\vspace{2mm}
\begin{align}\label{delta-Pi-NotFinal}
    \delta \Pi_R(P) &= \frac{g_s^2 t_{\mathrm{R}} T^2}{6\pi^2 \tau} \left[N_c + N_f(1-2^{-(\ell+1)})\right]\Gamma(\ell+3) \zeta(\ell+2) [I_\theta(1,0) - 3 I_\theta(1,2)], \quad \ell>-1
\end{align}
\vspace{2mm}
where $\zeta(\ell)$ is the Riemann Zeta, and the angular integrals, $I_\theta(1,0)$ and $I_\theta(1,2)$ are evaluated as,
\begin{subequations}
    \begin{align}
        &I_\theta(1,0) =  4 \left[ \frac{\left(p^0+i\epsilon\right)}{2p} \ln \left(\frac{p^0 + p + i\epsilon}{p^0 - p + i\epsilon}\right) - 1\right], \\
        &I_\theta (1,2) =  - \frac{2 \left(p^0 + i \epsilon\right)}{p^3} \left[p^2 - 2 \left(p^0 + i\epsilon\right)^2\right] \ln\left(\frac{p^0 + p + i \epsilon}{p^0 - p + i \epsilon}\right) - \frac{8(p^0+i\epsilon)^2}{p^2} + \frac{4}{3}.
    \end{align}
\end{subequations}
Taking the static limit of this non-equilibrium correction to self-energy, we can determine the correction to the Debye screening mass as,
\begin{align}
    \delta m_D^2 &= -\lim_{\substack{p^0 = 0,\, \bf{p} \to 0}} \delta \Pi_R(P).
\end{align}
In the static limit, the angular integrals, which are the only gluon momentum-dependent quantity in Eq.~(\ref{delta-Pi-NotFinal}), become,
\begin{subequations}
\begin{align}
    &\lim_{p^0=0,\,\,\bf{p}\rightarrow 0} I_\theta(1,0) = -4,\label{App20}\\
    &\lim_{p^0=0,\,\,\bf{p}\rightarrow 0} I_\theta(1,2) = \frac{4}{3}\label{App21}.
\end{align}
\end{subequations}
With this, we get the expression for the non-equilibrium correction to the screening mass squared as,
\begin{align}
    \delta m_D^2 = \frac{4 g_s^2 t_\mathrm{R}T^2}{3\pi^2 \tau} \left[N_c + N_f(1-2^{-(\ell+1)})\right]\Gamma(\ell+3) \zeta(\ell+2), \quad \quad \ell>-1.
    \label{eq:deltamD}
\end{align}
The total Debye screening mass, including the equilibrium part, in an expanding medium within the ERTA framework, can be written as,
\begin{align}
    m_D^2 = \frac{g_s^2 T^2}{6} (N_f + 2N_c) + \frac{4 g_s^2 t_\mathrm{R} T^2 }{3\pi^2 \tau} \left[N_c + N_f(1-2^{-(\ell+1)})\right] \Gamma(\ell+3) \zeta(\ell+2), \quad \quad \ell>-1.
    \label{eq:finalmD}
\end{align}
The modification of the screening mass can affect the real and imaginary part of quarkonia potential as discussed in section \ref{II.1}, which in turn can modify the in-medium properties of heavy quarkonia, which we discuss in the next section.

%%%%%%%%%%%%%%%%%%%%%%%%%%%%%%%%%%%%%%%%%%%%%%%%%%%%%%%%%%%
\subsection{In-Medium properties of quarkonium states}\label{II.3}
%%%%%%%%%%%%%%%%%%%%%%%%%%%%%%%%%%%%%%%%%%%%%%%%%%%%%%%%%

We now focus on two physical observables essential for understanding the behavior of heavy quarkonium states in a thermal medium: the thermal width, which quantifies the dissociation rate, and the binding energy. These parameters provide valuable insights into the stability and longevity of heavy quark-antiquark bound states such as charmonium and bottomonium in hot and dense environments, like those created in heavy-ion collisions.

To compute the binding energy, we use the real part of the in-medium potential, which includes both Coulomb and string contributions. The potential reduces to its vacuum form at short distances and approaches a screened Coulomb behavior at large separations ($\hat{r}>>1$). Following Ref.~\cite{Agotiya:2008ie}, the large $\hat{r}$ approximation leads to a coulombic potential given by,
\begin{align}
V(r)=-\frac{2\sigma}{m_D^2r}+\frac{2\sigma}{m_D}-\alpha m_D,\label{Vr}
\end{align}
Solving the Schr\"{o}dinger equation with this potential yields the quantized energy levels of quarkonium systems such as charmonium and bottomonium\footnote{In the high temperature limit, $\alpha m_D >> 2\sigma/m_D$ and hence the second term can be neglected in both Eqs.~\eqref{Vr} and \eqref{BE_TB}.},
\begin{eqnarray}
\label{bind1}
E_n=-\frac{1}{n^2}\frac{m_Q\sigma^2}{m_D^4}+\frac{2\sigma}{m_D}-\alpha m_D,
\label{BE_TB}
\end{eqnarray}
where $m_Q$ is the heavy quark mass and $m_D$ is the Debye screening mass, whose explicit form is given in Eq.~\eqref{eq:finalmD}. The ground ($n=1$) and first excited ($n=2$) states correspond to the $J/\psi$ and $\psi'$ for charmonium, and $\Upsilon$ and $\Upsilon'$ for bottomonium. In this work, we focus on the ground-state ($J/\psi$) binding energy given by $E_b=V(\infty)-E_1$; results for excited states can be obtained similarly. The full Schr\"{o}dinger equation with the complete in-medium potential will be solved numerically in a future study to obtain precise binding energies.

%%%%%%%%%%%%%%%%%%%%%%%%%%%%%%%%%%%%%%%%%%%%%%%%%%%%%%%%%%%%%%%%%%%

%\subsubsection{Thermal width of quarkonium states}

The thermal width  ($\Gamma$) provides the measure of the in-medium decay or dissociation probability of a bound quarkonium state and can be evaluated from the imaginary part of the in-medium potential between the heavy quark and antiquark. Assuming knowledge of the spatial wave function $\Psi({\bf r})$ of the quarkonium state, the thermal width $\Gamma$ can be expressed as an expectation value of the imaginary part of the potential $\text{Im}\,V(r,T)$, given in Eq.~\eqref{eq:ImV1}, evaluated over the wave function profile. Specifically, it is given by~\cite{Nilima:2022tmz,Thakur:2013nia,Thakur:2020ifi},
\begin{align}
    \Gamma(T) = - \int d^3{\bf r}\, \left|\Psi({\bf r})\right|^2\,\text{Im}\,V({\bf r};T)~, \label{Gamma}
\end{align}
where $T$ denotes the temperature. The imaginary part of the potential accounts for the Landau damping and in-medium scattering processes that lead to the dissociation of the bound state. This formulation is derived using first-order quantum mechanical perturbation theory, where the unperturbed wave functions are taken to be the Coulombic eigenfunctions of the bound state system.

For our analysis, we consider the ground states of the charmonium system, corresponding to $J/\psi$. The radial parts of the Coulombic wave function for the $1s$ state is given by,
\begin{align}
\Psi_{1s}(r) &= \frac{1}{\sqrt{\pi r_B^3}}\,e^{-r/r_B},
\label{psi}
\end{align}
where $r_B = \dfrac{m_D^2}{m_Q \sigma}$ is the Bohr radius of the quarkonium system with $\sigma$ as the string tension. The Bohr radius reflects the
spatial extent of the bound state and plays a 
crucial role in determining the overlap with the 
imaginary potential. Since the wave function 
$\Psi_{1s}(r)$ is spherically symmetric, the 
integral in Eq.~\eqref{Gamma} can be simplified 
to a radial form. The magnitude of $\Gamma$ thus 
captures how the in-medium effects, encoded in $\text{Im}\,V(r,T)$, modify the stability of the quarkonium states.
We note that this approach provides a good estimate for the ground-state width, while a more detailed calculation using the full numerically obtained wave function can be considered in future work.

%-------------------------------------------
\section{Results and discussions}\label{III}
%-------------------------------------------

\begin{figure}[tbh]
    \begin{center}
        \includegraphics[scale=0.7]{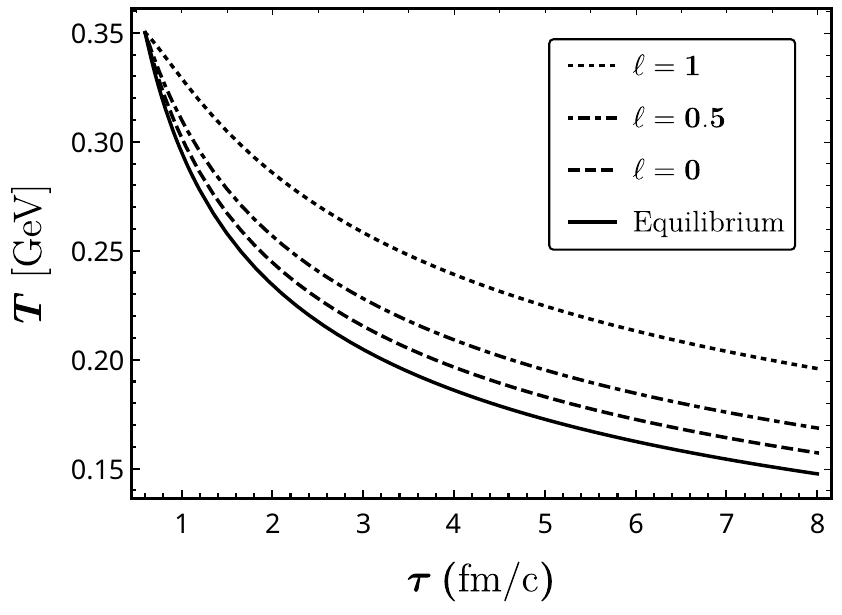}
        \caption{Temperature evolution of the medium as a function of its proper time ($\tau$) for equilibrium and non-equilibrium scenarios within the ERTA for different values of the momentum-dependence parameter $\ell$.}
        \label{T_vs_tau}
    \end{center}
\end{figure}
To study the evolution of quarkonia and their properties, 1-D boost-invariant longitudinal expansion of the medium has been considered. The temperature evolution of the expanding fluid can be obtained from the hydrodynamical equations. From the ERTA-based hydrodynamics, the temperature evolution in the massless limit as a function of the proper time $\tau$, was obtained by numerically solving Eq.~(\ref{T-evo}). The shear viscosity $\eta$ depends on the momentum-dependence parameter $\ell$ as $\eta = 2 t_R\beta^{\ell+1} \big[N_c N_f J_{3+\ell,2}^{(q)} + \left(N_c^2 - 1\right) J_{3+\ell,2}^{(g)}\big]$, where the superscripts $(q/g)$ represents the quark or gluonic distribution function within the thermodynamic integral.

For quantitative estimation, we choose the initial temperature  $T=350$ MeV at initial proper time $\tau_0=0.6$ fm/c and the momentum-independent part of the relaxation time is taken to be inversely proportional to the temperature T in the medium, $\tau_R=\kappa/T$, where $\kappa=0.125$  \cite{PHENIX:2008uif}. The temperature profile for the longitudinal expansion is depicted in Fig.~\ref{T_vs_tau}. Notably, the momentum dependence of the relaxation time has a substantial impact on the temperature evolution.  The black solid curve corresponds to the ideal hydrodynamic (equilibrium) case, where the system remains locally equilibrated and the temperature decreases most rapidly with time following the standard scaling law $T(\tau) \propto \tau^{-1/3}$. The dashed curves represent non-equilibrium scenarios with different choices of $\ell$: the standard RTA case with $\ell = 0$, along with the values $\ell = 0.5$ and $\ell = 1$. From Eq.~\eqref{tau_R-param}, it is understood that the relaxation time becomes more sensitive to the particle momenta as the value of $\ell$ increases. A larger $\ell$ implies that the high-momentum particles take longer to equilibrate. Consequently, the effective thermalization of the medium slows down, and the temperature drops more slowly compared to the equilibrium case and the standard RTA estimation.

\begin{figure}[tbh]
\begin{center}
\includegraphics[scale=0.525]{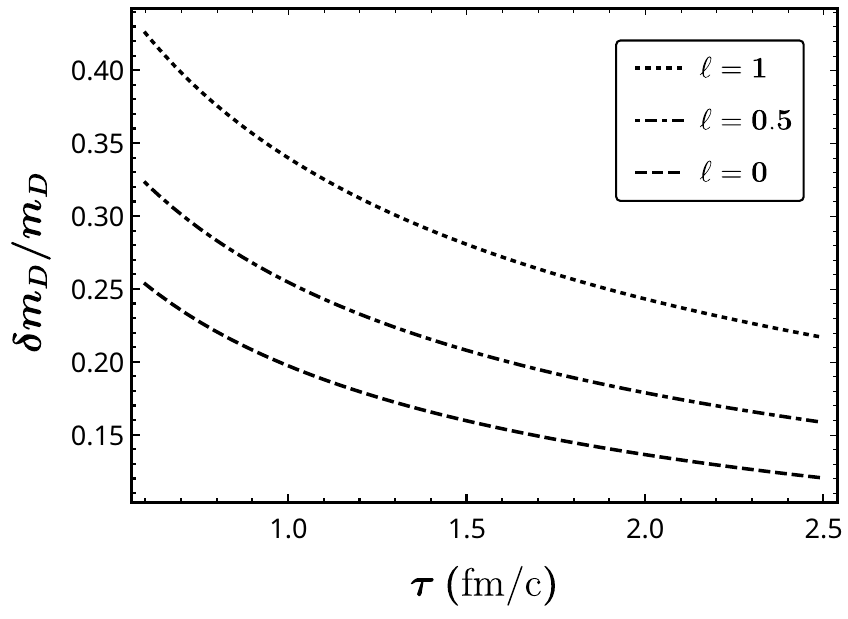}
\includegraphics[scale=0.510]{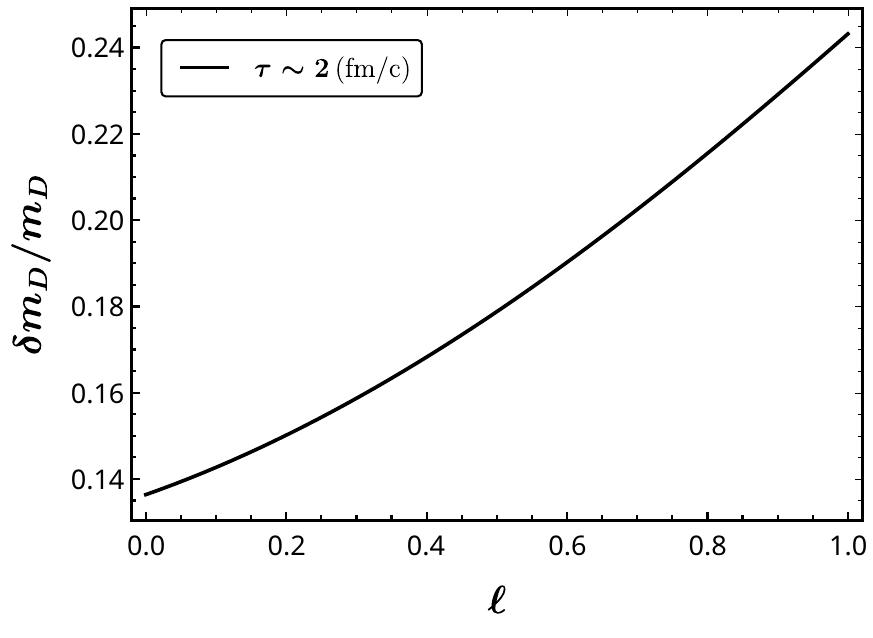}
\caption{ \small The ratio of $\delta m_D$ to the total screening mass $m_D$ is plotted against the proper time $\tau$ for different values of $\ell$ (left panel). In the right panel, this ratio is plotted against $\ell$ at $\tau= 2$ fm/c which corresponds to $T\sim 235$ MeV in the equilibrium scenario.}
\label{screening_mass_ratios}
\end{center}
\end{figure}

Fig.~\ref{screening_mass_ratios} (left panel) illustrates the significance of non-equilibrium correction to the screening mass. This is achieved by plotting the proper time evolution of the ratio $\dfrac{\delta m_D}{m_D}$. The non-equilibrium correction $\delta m_D$ is obtained within the ERTA framework as described in Eq.~\eqref{eq:deltamD}. At early times, when the system is hotter and farther from equilibrium, the dissipative corrections to the screening mass are more prominent. As the system evolves and cools with time, these corrections diminish. We note that the ERTA framework yields significantly larger corrections compared to the conventional RTA result ($\ell = 0$) throughout the entire evolution. The dependence of $\ell$ on the ratio $\dfrac{\delta m_D}{m_D}$ for a fixed proper time ($\tau = 2~\mathrm{fm}/c$) is shown in Fig.~\ref{screening_mass_ratios} (right panel). We observe that the magnitude of $\dfrac{\delta m_D}{m_D}$ increases with $\ell$, indicating that systems with stronger momentum dependence in their relaxation dynamics exhibit greater deviations from equilibrium. This observation underscores the crucial role of microscopic relaxation dynamics in determining in-medium properties, which, in turn, can affect the quarkonia potential in the QCD medium.
%Can't we make a comment about the fact that the negative part of the potential is affected more as compared to the positive part?

\begin{figure}
\begin{center}
\includegraphics[scale=0.516]{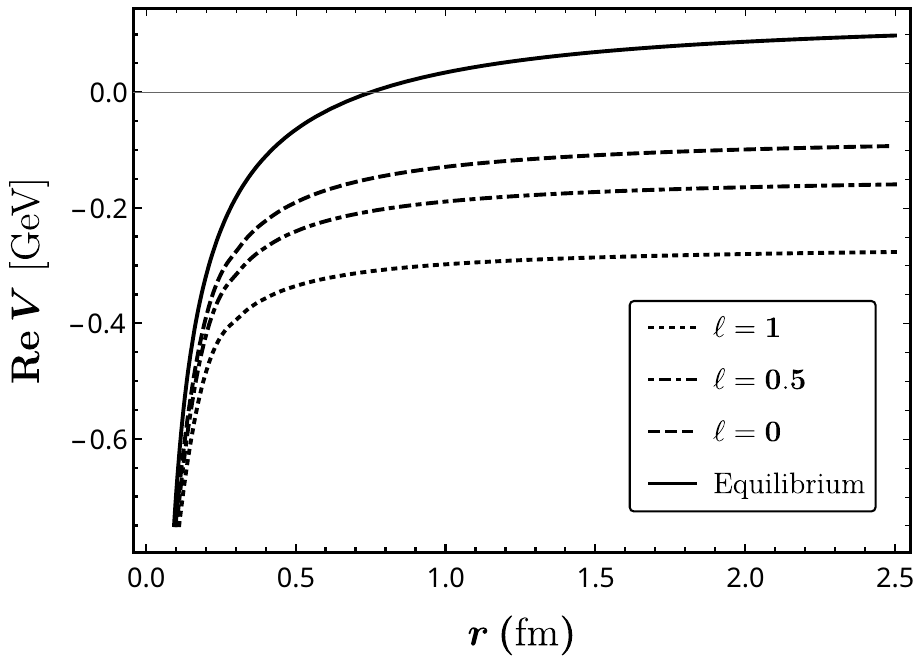}
\includegraphics[scale=0.500]{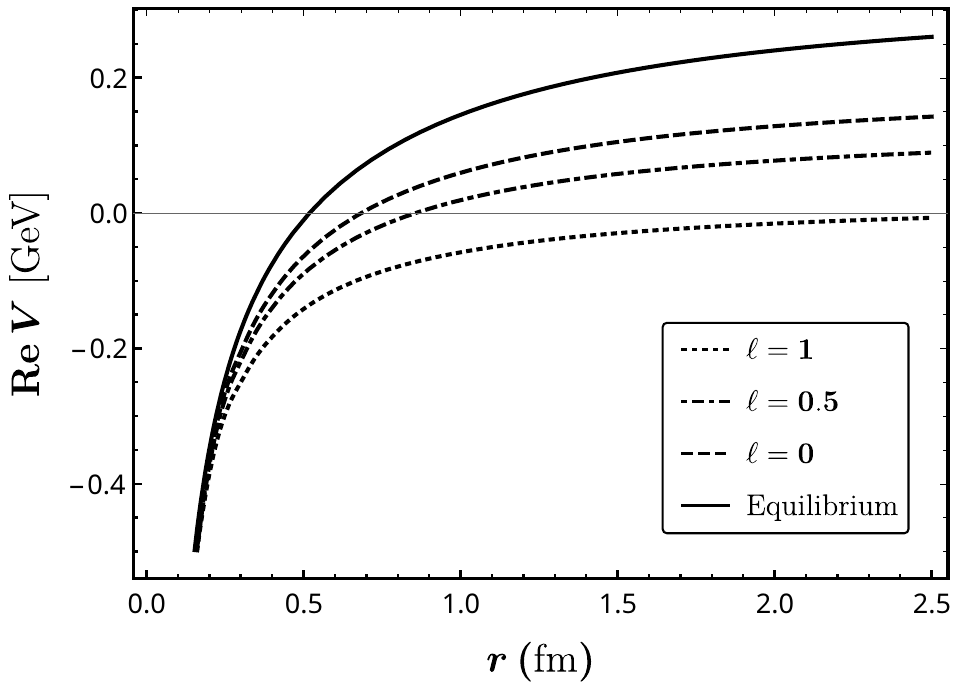}
\caption{ \small The real part of a quarkonia potential is plotted against the distance, $r$ between the heavy quarks, for various values of $\ell$ at proper times, $\tau=0.6$ fm/c corresponding to $T \sim 350$ MeV (left panel) and $\tau=2$ fm/c corresponding to $T \sim 235$ MeV (right panel).}
\label{ReV_plotss}
\end{center}
\end{figure}

Fig.~\ref{ReV_plotss} depicts the real part of the in-medium quarkonium potential as a function of the distance $r$ between the heavy quark and anti-heavy quark, for different values of the momentum-dependence parameter $\ell$. The left panel corresponds to an early proper time $\tau = 0.6~\mathrm{fm}/c$, where the ideal system is hotter ($T \sim 350~\mathrm{MeV}$), and the right panel corresponds to a later time $\tau = 2~\mathrm{fm}/c$, where the temperature has dropped to $T \sim 235~\mathrm{MeV}$. In both panels, the black solid curve represents the equilibrium potential, and $\ell = 0$ case represents the non-equilibrium scenario within the RTA approximation. The non-equilibrium corrections have a crucial role, especially at larger values of $r$. The ERTA framework introduces substantial modifications to the RTA-based corrections. As the value of $\ell$ increases, the potential exhibits enhanced screening, indicating stronger medium modifications driven by non-equilibrium effects. The potential flattens more rapidly with increasing distance for higher values of $\ell$, which can potentially reduce the binding strength of heavy quark pairs at larger separations. While the qualitative behavior of the real part of the potential in the later stage of evolution ($\tau = 2~\mathrm{fm}/c$) remains the same in comparison with the early time ($\tau = 0.6~\mathrm{fm}/c$), the magnitude differs. This difference arises due to temperature-dependent screening in the medium. Notably, we observe that the impact of dissipative corrections is more pronounced in the early stages of evolution.

\begin{figure}
\begin{center}
\includegraphics[scale=0.5150]{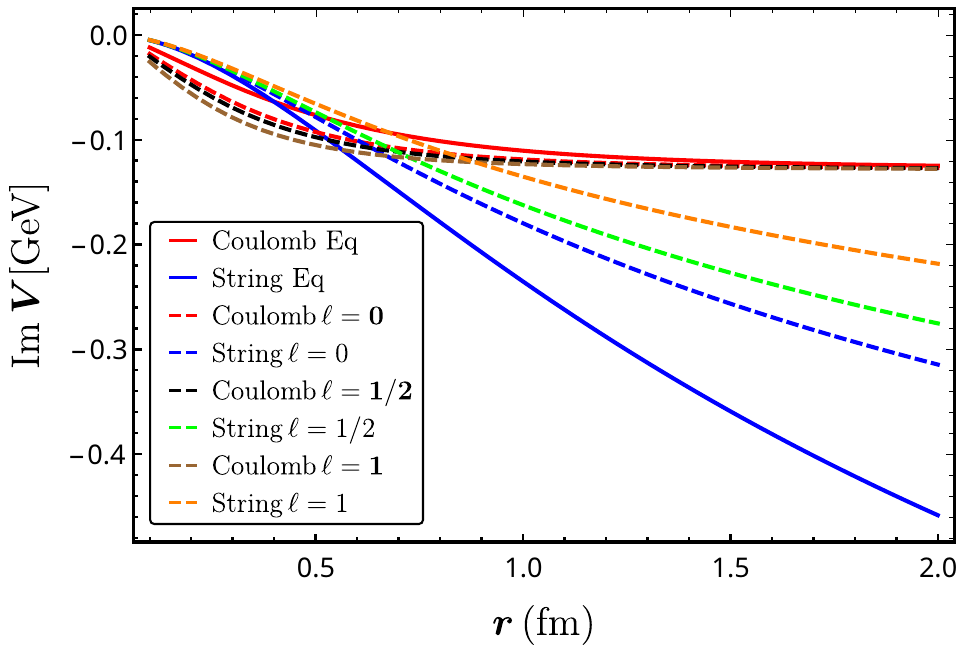}
\includegraphics[scale=0.500]{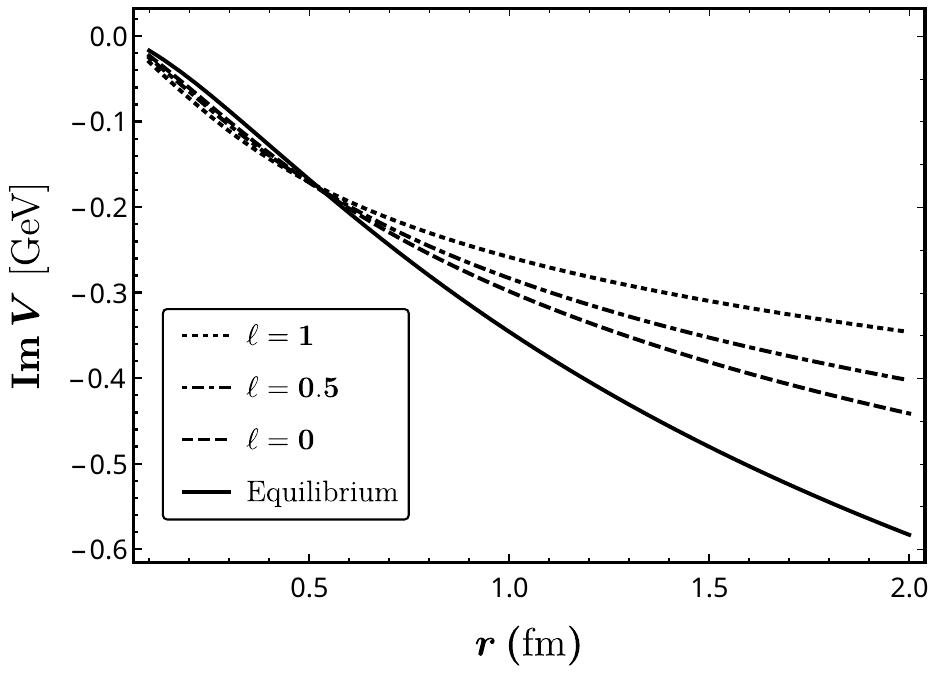}
\caption{ \small The imaginary part of a quarkonia potential is plotted against the distance $r$ between the quark anti-quark pair for various values of $\ell$ at $\tau=0.6$ fm/c. The left panel shows both the contributions from coulomb and string parts to the imaginary potential for each $\ell$ along with the equilibrium case, and the right panel shows the total value of the imaginary potential for both the equilibrium and non-equilibrium scenarios.}
\label{ImV_plotss_6}
\end{center}
\end{figure}

\begin{figure}
\begin{center}
\includegraphics[scale=0.510]{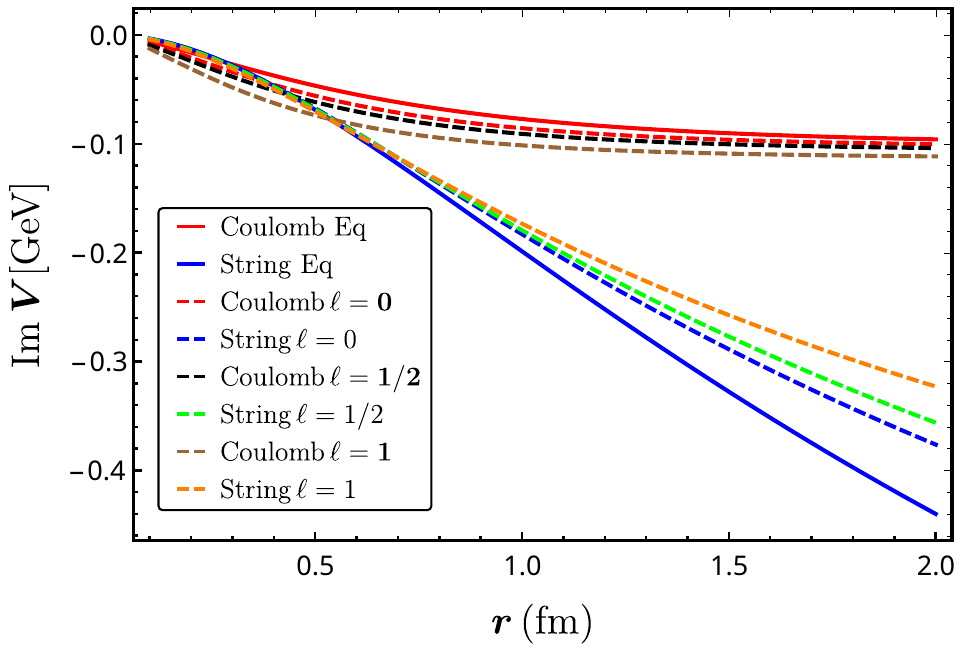}
\includegraphics[scale=0.510]{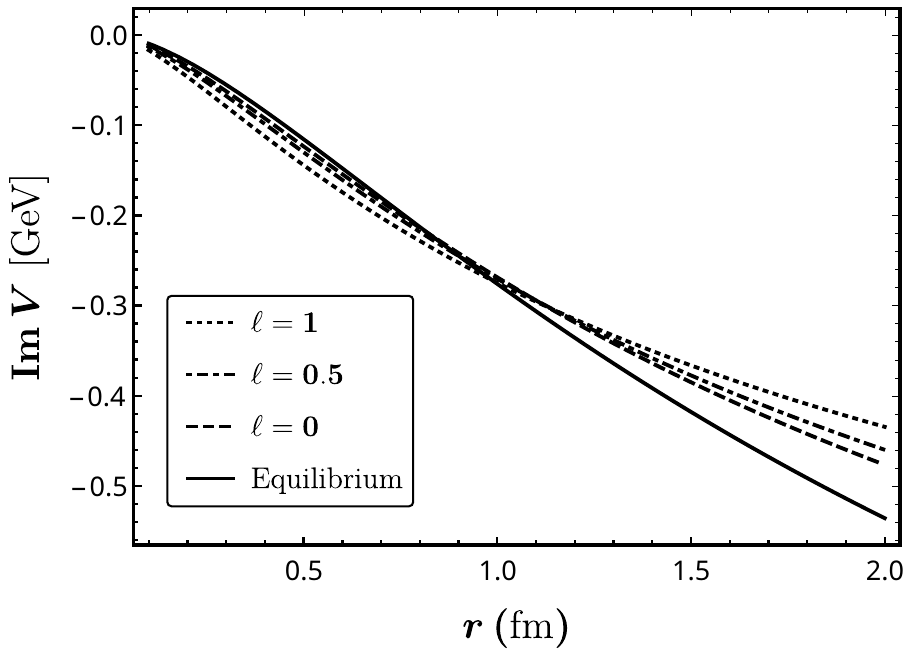}
\caption{ \small The imaginary part of a quarkonia potential is plotted against the distance $r$ between the quark antiquark pair for various values of $\ell$ at $\tau=2$ fm/c. The left panel shows both the contributions from coulomb and string parts to the imaginary potential for each $\ell$ along with the equilibrium case, and the right panel shows the total value of the imaginary potential for both the equilibrium and non-equilibrium scenarios.}
\label{ImV_plotss2}
\end{center}
\end{figure}

The imaginary part of the in-medium quarkonium potential as a function of the inter-quark distance $r$ within the ERTA framework at $\tau=0.6$ fm/c is plotted in Fig.~\ref{ImV_plotss_6}. The left panel separately depicts the Coulomb and String contributions to the imaginary part of the potential, which are the first and second term of Eq.~(\ref{eq:ImV1}), respectively, for both the equilibrium and non-equilibrium scenarios, while the right panel displays the total imaginary potential. We observe that the string part shows considerable sensitivity to the momentum dependence of the collision dynamics in the medium, whereas the Coulomb part shows a weaker dependence on the same as compared to the string part. Fig.~\ref{ImV_plotss2} illustrates the behavior of the imaginary part of the potential at a later time, $\tau=2$ fm/c. In contrast to the late-time scenario, dissipative corrections have a significant impact on the imaginary part of the potential during the early stages of evolution. These effects become increasingly pronounced with larger values of $\ell$, suggesting that momentum-dependent microscopic interactions at early times are important. Although this effect is less prominent at later times ($\tau = 2~\mathrm{fm}/c$), as shown in Fig.~\ref{ImV_plotss2}, it remains non-negligible. 
The total imaginary part of the potential exhibits a narrowing deviation between the $\ell$-dependent curves and the equilibrium case over time. These findings indicate that while dissipative effects gradually weaken as the system evolves, they remain non-negligible even at $\tau = 2$~fm/c, particularly for larger values of $\ell$.

% Next, we integrate out the dependence on the heavy quark-antiquark separation distance $r$ in both the real and imaginary parts of the potential by employing a Coulomb-type probability distribution function as in Eq.~\eqref{psi}. This allows us to compute the binding energy by using Eq.~\eqref{BE_TB}, and to evaluate the thermal width or dissociation probability $\Gamma_{1s}$ via Eq.~\eqref{Gamma}.

\begin{figure}
\begin{center}
\includegraphics[scale=0.510]{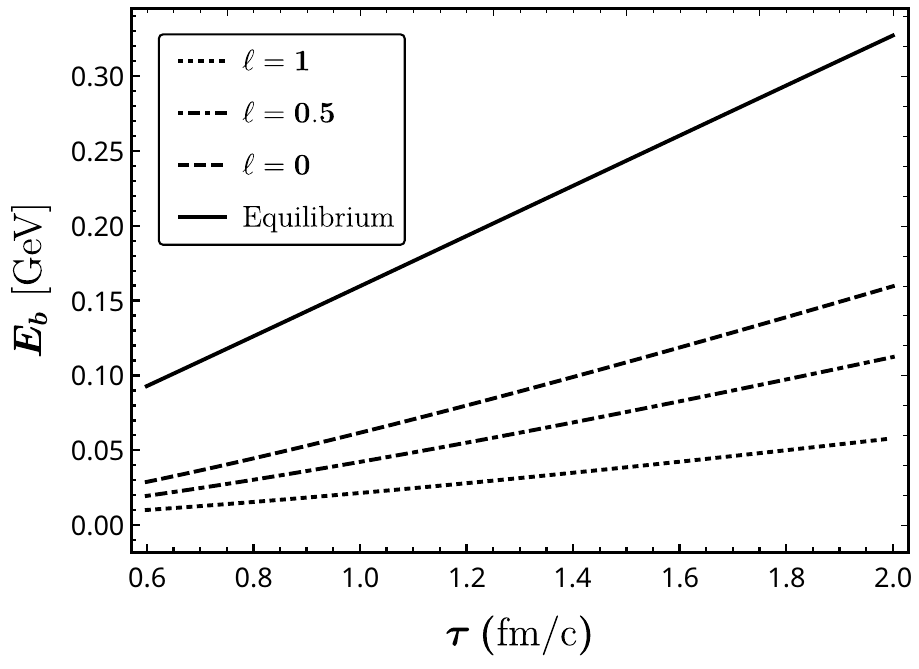}
\includegraphics[scale=0.442]{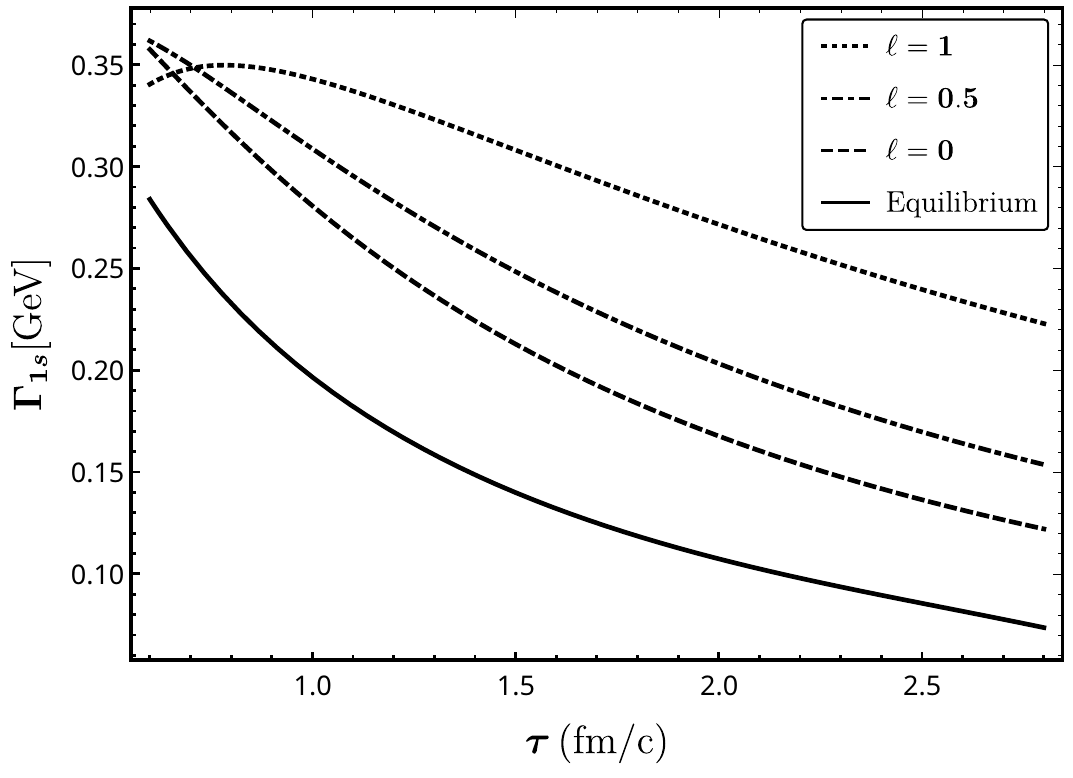}
\caption{ \small The binding energy of the $J/\Psi$ ground state as a function of proper time, $\tau$ at various values of $\ell$ (left panel). The line width at various values of the momentum dependence parameter $\ell$ (right panel).}
\label{Binding_width}
\end{center}
\end{figure}

Next, following section \ref{II.3}, we focus on the ground state of $J/\psi$ quarkonium in the present study. This state, often modeled as a $1s$ bound state of a charm–anticharm $(c\bar c)$ pair is of particular interest due to its relatively large binding energy and stability in the QGP. For quantitative estimation, we take the charmonium mass as $ m_c = 1.25$ GeV. Fig.~\ref{Binding_width} depicts the binding energies (left panel) and thermal widths (right panel) as a function of proper time $\tau$ within the ERTA framework. In both plots, the black solid line represents the equilibrium result, corresponding to the case without any viscous corrections. The binding energy is found to increase with proper time, reflecting its inverse temperature dependence (Fig.~\ref{Binding_width}, left panel). As the system evolves and cools, the quarkonium states become more strongly bound due to reduced thermal screening in the medium. We also note that the inclusion of dissipative effects leads to a reduction in binding energy which can be explained due to the fact that the viscous corrections enhance the screening of the heavy-quark potential and as a result, the interaction between the quark and antiquark weakens. Furthermore, as the momentum dependence parameter $\ell$ increases in value, the screening effect becomes more pronounced, resulting in a further decrease in the binding energy. It is also worth noting that the ERTA framework leads to notable modifications compared to the standard RTA corrections ($\ell=0$ case).

The decay width of the $1s$ bound state was determined as a function of the proper time $\tau$ using Eq.~(\ref{Gamma}) and the Bohr radius defined after Eq.~(\ref{psi}). We notice that throughout the evolution of the QGP medium, dissipative effects tend to increase the thermal width of the quarkonium state compared to the case when equilibrium is assumed (Fig.~\ref{Binding_width}, right panel). This shows that dissipative effects decreases the lifetime of the ground state, which implies that the dissociation rates increases and hence it should lead to a stronger quarkonia suppression. We also note that this increase in thermal width compared to the equilibrium scenario increases with the value of $\ell$. These results are consistent with Fig.~\ref{screening_mass_ratios} (right panel), which shows that higher values of $\ell$ leads to greater screening in the medium which should weaken the inter-quark potential and hence should lead to a shorter lifetime of the quarkonia state, and in turn the thermal width should increase. Hence, we find that the momentum dependence in the relaxation time via the parameter $\ell$ has a crucial role in determining the behavior of the thermal width. As the system cools down with increasing proper time $\tau$, the temperature decreases, leading to a reduction in the number of thermally excited quarks and gluons as well as the screening mass in both the equilibrium and non-equilibrium scenarios as shown in Fig.~\ref{screening_mass_ratios} (left panel). Consequently, we see in Fig.~\ref{Binding_width} (right panel) that the decay width naturally becomes smaller as the system evolves since the inter-quark potential strengthens and the QGP becomes less capable of dissociating the quarkonium state. Apart from this, we observe that the $\ell=1$ case shows a maximum around $\tau=0.8$ fm/c. This might be due to an interplay of the screening and the number of excited quarks and gluons available to cause quarkonia dissociation, which increases the thermal width momentarily. This non-monotonous behavior should be investigated in a future study. 

\section{Summary and outlook}\label{IV}
%####################################################
In this study, we explored the properties of quarkonia in a QCD medium within a newly proposed microscopic theory that incorporates particle-momentum dependence in the relaxation time. We begin by estimating the non-equilibrium corrections to the longitudinal component of the gluon self-energy within the one-loop HTL approximation. The non-equilibrium dynamics of the QCD medium are modeled within the ERTA-based modified kinetic theory. We analyzed the sensitivity of quarkonia potential to non-equilibrium effects arising from the expansion of the medium. The in-medium modifications to the potential are incorporated through the dielectric response of the QCD medium. To quantify these effects, we analyzed the binding energy and thermal width of quarkonia states. We observe that the bulk viscous corrections are minimal, whereas shear viscous corrections are non-negligible and significantly affect the binding energy and thermal width. We compared the ERTA-based calculations of the system of massless particles with those obtained from standard RTA results. Notably, we observe that the momentum dependence of the relaxation time has a sizable impact on the non-equilibrium corrections to the physical observables. Our findings highlight that incorporating the specifics of microscopic interactions on the collision timescale is crucial for the study of quarkonia in the QCD medium. This dependence has potential implications for phenomenological studies as well.

This work represents a step towards the realistic description of the quarkonia potential in the QCD medium. For a more realistic study, two aspects need further attention. First, it is necessary to incorporate effective modeling of thermal medium effects within the ERTA framework, which includes a consideration of the temperature dependence of the $\ell$-parameter \cite{Mukherjee:2025dqp}. Although this extension poses mathematical challenges, it may reveal non-trivial contributions of bulk viscous corrections. Second, we need to consider a realistic 1+3D expansion of the QGP medium, where the medium may evolve according to the Israel-Stewart-like second-order hydrodynamic theories. We intend to explore these aspects in the near future.

\section*{Acknowledgments}
We acknowledge Vinod Chandra, Amaresh Jaiswal and Santosh Kumar Das for the useful discussions.  This research of R.G. was funded in part by the U.S. National Science Foundation under Grant No.~PHY-2209470 and PHY-2514933. M.K. acknowledges the Department of Science and Technology (DST), Govt. of India, for the INSPIRE-Faculty award (DST/INSPIRE/04/2024/001794), and the Faculty Research Scheme (FRS project number: MISC 0240) at IIT (ISM) Dhanbad, India.

\appendix
%%%
\section{First-order correction to distribution function $\delta f_{(1)}$ in Bjorken expansion}
\label{Appendix:B}
Here, we show that the form of $\delta f_{(1)}$ given in Eq.~(\ref{del_f-ERTA}) can be reduced to Eq.~(\ref{shearf}) in a longitudinally expanding fluid that is boost-invariant. For this, we need to calculate the form of $K^\mu K^\nu \sigma_{\mu\nu}$ in the local rest frame of the fluid using the cartesian coordinates to be able to express it in terms of $k$ and the angle between the gluon and the particle momenta, $\theta_{kp}$. We begin with the components of the shear-velocity tensor, $\sigma^{\mu\nu}$ in cartesian co-ordinates obtained by differentiating the fluid velocity $U^\mu=(t/\tau,0,0,z/\tau)$ using its definition, $\sigma^{\mu\nu}=\Delta^{\mu\nu}_{\alpha\beta} \partial^\alpha U^\beta$,
\begin{align}
\sigma^{\mu\nu} =
\begin{pmatrix}
-\dfrac{2z^2}{3\tau^3} & 0 & 0 & -\dfrac{2zt}{3\tau^3} \\
0 & \dfrac{1}{3\tau} & 0 & 0 \\
0 & 0 & \dfrac{1}{3\tau} & 0 \\
-\dfrac{2zt}{3\tau^3}  & 0 & 0 & -\dfrac{2t^2}{3\tau^3}
\end{pmatrix}
\quad \text{in Cartesian coordinates } (t, x, y, z).
\end{align}
Boosting to the local rest frame of the fluid via a Lorentz transformation $\Lambda^\alpha_{\,\,\beta}(\eta)$ where $\eta=\tanh^{-1}(z/t)$ is the space-time rapidity, the components of the $\sigma^{\mu\nu}_{\text{LRF}}$ tensor become,
\begin{align}
\sigma^{\mu\nu}_{\text{LRF}} =
\begin{pmatrix}
0 & 0 & 0 & 0 \\
0 & \dfrac{1}{3\tau} & 0 & 0 \\
0 & 0 & \dfrac{1}{3\tau} & 0 \\
0 & 0 & 0 & -\dfrac{2}{3\tau}
\end{pmatrix}
\quad \text{in Cartesian coordinates } (t, x, y, z).
\end{align}
With this, we have,
\begin{align}
    k_\mu k_\nu \sigma_\text{LRF}^{\mu\nu} &= k_x^2 \sigma_\text{LRF}^{11} +  k_y^2 \sigma_\text{LRF}^{22} +  k_z^2 \sigma_\text{LRF}^{33},
\end{align}
where $K_\mu = (E_k, -k_x,-k_y,-k_z)$ are the components of particle momenta in the cartesian coordinates in the local rest frame of the fluid. Hence, we can express them as $k_x=k \sin \theta_{kp} \cos \phi$, $k_y= k \sin \theta_{kp} \sin \phi$ and $k_z= k \cos \theta_{kp}$. This leads to,
\begin{align}
    K_\mu K_\nu \sigma_\text{LRF}^{\mu\nu}&=\frac{1}{3\tau}k^2 \sin^2\theta_{kp}-\frac{2}{3\tau}k^2 \cos^2\theta_{kp} \nonumber\\
    &=\frac{k^2}{3\tau}(1-3 \cos^2 \theta_{kp}).
\end{align}
With this, we can write the form of the non-equilibrium correction to the distribution function as given in Eq.~(\ref{del_f-ERTA}).
\section{Estimation of $I_\ell$  integrals}\label{Appendix2}
We consider the form for the equilibrium distribution function,
\begin{align}
    &f_0 = \frac{1}{e^{k/T}+a}, && f_0 \Tilde{f}_0 = \frac{e^{k/T}}{\left(e^{k/T}+a\right)^2},
\end{align}
where $a=1,0,-1$ for particles following Fermi-Dirac, Maxwell-Boltzmann, or Bose-Einstein statistics, respectively, and $\Tilde{f}_0 = 1 -a f_0$. The $I_\ell$ integral can be defined as,
\begin{align}
    I_\ell= T^{\ell+3} \int^\infty_0 {dz} \, \frac{z^{\ell+2}e^z}{\left(e^z+a\right)^2},
\end{align}
where we have substituted $k/T=z$ in Eq.~(\ref{I-gluon}). By defining $u= z^{\ell+2}$ and $v=-\frac{1}{e^z+a}$ and using integration by parts, we obtain
\begin{align}
    \frac{I_\ell}{T^{\ell+3}} &= -\int^\infty_0 v \frac{du}{dz} + uv|^\infty_0 \nonumber\\
    &= (\ell+2)\int^\infty_0 {dz}\, \frac{z^{\ell+1}}{e^z+a} + \lim_{z\rightarrow 0}\frac{z^{\ell+2}}{e^z+a} -\lim_{z\rightarrow \infty} \frac{z^{\ell+2}}{e^z+a}.
\end{align}
The limits in the above expression become $0$ since the exponential in the denominator increases faster than the polynomial in the numerator as $z \rightarrow \infty$. With this, the integral becomes,
\begin{align}
    I_\ell= T^{\ell+3}(\ell+2)\int^\infty_0 {dz}\, \frac{z^{\ell+1}}{e^z+a}.
\end{align}
This is a standard integral which can be expressed in terms of the Dirichlet $\eta(\ell)$ function or Riemann $\zeta(s)$ function, based on if $a=1$ for FD statistics (quarks) or if $a=-1$ for BE statistics (gluons).
\subsubsection{ Quark sector}
%\subsection{Dirchlet $\eta(s)$ for Quark sector}
The Dirichlet $\eta(s)$ function is defined as,
\begin{align}
    \eta(s) = \sum^\infty_{n=0} \frac{(-1)^{n-1}}{n^s} = \frac{1}{1^s} - \frac{1}{2^s} + \frac{1}{3^s} - \dots
\end{align}
This is convergent for $s>0$ and it is related to the Riemann Zeta function, $\zeta(s)$ by,
\begin{align} \label{etaToZeta}
    \eta(s) = (1-2^{(1-s)})\zeta(s).
\end{align}
At $s=1$, the simple pole for $\zeta(s)$ is canceled by the zero in the other factor, leading to a finite value of $\ln 2$. Integrals like the following can be expressed in terms of the $\eta(s)$ functions,
\begin{align}
    \int^\infty_0 dz \frac{z^{s-1}e^z}{e^z+1}=\eta(s)\Gamma(s).
\end{align}
Using this, we can write $I_{\ell}^{(q)}$ as,
\begin{align}
    I_{\ell}^{(q)} = T^{\ell+3}(\ell+2)\Gamma(\ell+2)\eta(\ell+2), \quad \quad \ell > -2
\end{align}
Using the property $\Gamma(s+1)=s\Gamma(s)$ and Eq.~\eqref{etaToZeta}, we have,
\begin{align}
    I_{\ell}^{(q)} = T^{\ell+3}\Gamma(\ell+3) (1- 2^{-(\ell+1)})\zeta(\ell+2), \quad \quad \ell>-2
\end{align}
\subsubsection{ Gluon sector}
%\subsubsection{Riemann $\zeta(s)$ for Gluon sector}
The Riemann zeta function is defined as,
\begin{align}
    \zeta(s) = \sum^\infty_{n=0} \frac{1}{n^s} = \frac{1}{1^s} + \frac{1}{2^s} + \frac{1}{3^s} + \dots
\end{align}
This series is convergent for $s>1$. The following form of integrals can be expressed in terms of the $\zeta(s)$ functions,
\begin{align}
    \int^\infty_0 dz \frac{z^{s-1}e^z}{e^z-1}=\zeta(s)\Gamma(s).
\end{align}
This leads to the form of $I_{\ell}^{(g)}$ as,
\begin{align}
I_{\ell}^{(g)} 
%&= T^{\ell+3}(\ell+2)\Gamma(\ell+2)\zeta(\ell+2) \nonumber\\
&= T^{\ell+3} \Gamma(\ell+3) \zeta(\ell+2) ,\quad \quad \ell>-1
\end{align}
Hence, for values of the momentum dependence parameter $\ell>-1$ (which is within our range of interest), we get the following expression for the building block of the gluon self-energy,
\begin{align}
    N_f I_{\ell}^{(q)}+ N_c I_{\ell}^{(g)} = \left[N_c + N_f(1-2^{-(\ell+1)})\right]T^{\ell+3}\Gamma(\ell+3) \zeta(\ell+2).
\end{align}

\bibliography{ref}
%############################################################################################################
\end{document}